# Astro2020 Science White Paper

# Stars at High Spatial Resolution

**Thematic Areas:**  ☐ Planetary Systems  ☒ Star and Planet Formation
☐ Formation and Evolution of Compact Objects  ☐ Cosmology and Fundamental Physics
☒ Stars and Stellar Evolution  ☐ Resolved Stellar Populations and their Environments
☐ Galaxy Evolution  ☐ Multi-Messenger Astronomy and Astrophysics

**Principal Author:**
Name: Kenneth G. Carpenter
Institution: NASA's Goddard Space Flight Center
Email: Kenneth.G.Carpenter@nasa.gov
Phone: 301-286-3453

**Co-authors:**

Gerard van Belle (Lowell Obs.), Alexander Brown (UCB), Steven R. Cranmer (UCB), Jeremy Drake (CfA), Andrea K. Dupree (CfA), Michelle Creech-Eakman (MRO), Nancy R. Evans (CfA), Carol A. Grady (Eureka Scientific), Edward F. Guinan (Villanova U.), Graham Harper (UCB), Margarita Karovska (CfA), Katrien Kolenberg (KU Leuven, UAntwerpen), Antoine Labeyrie (College de France), Jeffrey Linsky (UCB), Geraldine J. Peters (USC), Gioia Rau (NASA's GSFC/CUA), Stephen Ridgway (NOAO), Rachael M. Roettenbacher (Yale U.), Steven H. Saar (CfA), and Frederick M. Walter (SUNY), Brian Wood (NRL)

**Abstract:**
We summarize some of the compelling new scientific opportunities for understanding stars and stellar systems that can be enabled by sub-milliarcsec (sub-mas) angular resolution, UV/Optical spectral imaging observations, which can reveal the details of the many dynamic processes (e.g., evolving magnetic fields, accretion, convection, shocks, pulsations, winds, and jets) that affect stellar formation, structure, and evolution. These observations can only be provided by long-baseline interferometers or sparse aperture telescopes in space, since the aperture diameters required are in excess of 500 m – a regime in which monolithic or segmented designs are not and will not be feasible - and since they require observations at wavelengths (UV) not accessible from the ground. Such observational capabilities would enable tremendous gains in our understanding of the individual stars and stellar systems that are the building blocks of our Universe and which serve as the hosts for life throughout the Cosmos.

*Introduction*

*Understanding the formation, structure, and evolution of stars and stellar systems remains one of the most basic pursuits of astronomical science, and is a prerequisite to obtaining an understanding of the Universe as a whole. Furthermore, the existence and properties of exoplanet atmospheres depend on stellar UV, EUV, and X-ray emission and winds. The question of exoplanet habitability, which has great importance to the astronomical community and the general public, can only be addressed when we fully understand the present and past emission from host stars.* The evolution of structure and transport of matter within, from, and between stars are controlled by dynamic processes, such as variable magnetic fields, accretion, convection, shocks, pulsations, and winds. Future Long-Baseline (0.5-1.0 km) Space Interferometers (*LBSI*) will achieve resolutions of 0.1 milli-arcsec (mas), a gain in spatial resolution comparable to the leap from Galileo to HST. As a result, spectral imaging observations from such facilities will enable a quantum leap in our understanding of stars and stellar systems. In this whitepaper, we discuss the compelling new scientific opportunities for understanding the formation, structure, and evolution of stars and stellar systems that can be enabled by dramatic increases in UV-Optical angular resolution to the sub-mas level. A space-based UV-optical interferometer can provide direct spectral imaging of spatial structures and dynamical processes in the various stages of stellar evolution (e.g., Fig. 1) for a broad range of stellar types[8,22].

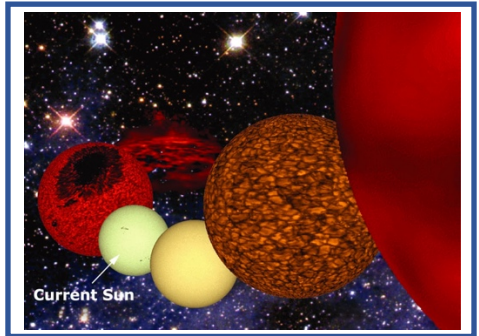

Fig. 1: Evolution of the Sun in time (left to right).

We discuss below the opportunities available for substantially improved observations and understanding of: young stellar systems; hot star rotation, disks, and winds; stellar pulsation across the HR-diagram and its impact on stellar structure and mass loss; convection in cool, evolved giant and supergiant stars; novae and supernovae. Fig. 2 (left) shows the resolution needed to resolve a wide variety of objects, (in minutes to hours with a 30-element *LBSI*), as a function of their distance and intrinsic size and (right) the time needed between images (just weeks-months!) to detect motions in many objects (e.g., mass transfer in binaries, pulsation driven surface brightness variations and convective cell structure in giants and supergiants, jet formation and propagation and changes in debris disks in young planetary systems due to orbiting planets, non-radial pulsations in and winds from stars, and the structure, evolution, and interaction with the ISM of nearby novae and super-novae.

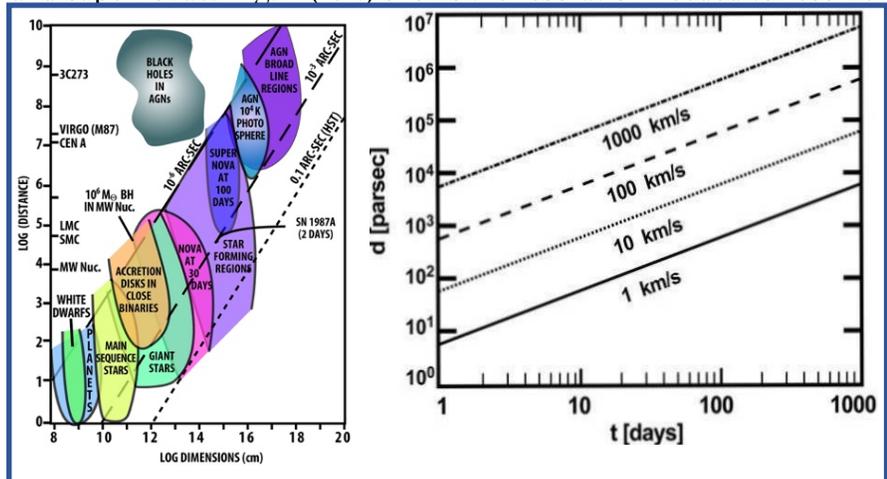

Fig. 2: **Left**: Resolution vs. size and distance (in pc) of object - a 500m *LBSI* could get below the 0.1mas line; **Right**: Minimum time interval between images required to resolve the motion of a feature moving at different speeds, as a function of the object's distance.

*Dynamic Processes in Young Stellar Systems: Star Formation, Protoplanetary Disks and Jets*

Protoplanetary disks are where the materials that can ultimately produce life-bearing worlds are assembled. For our own Solar System, the first 50 Myr span the formation and evolution of the proto-Solar nebula, the assembly of the meteorite parent bodies, the formation of the proto-Earth and proto-Mars, and the early phases of Heavy Bombardment. *If we are to comprehend how the history of our Solar System and planetary systems in general develop we need to understand the disks, how long they last, how they interact with their central stars, and how they evolve.*

For the first few million years, both young solar-type (T Tauri) and intermediate-mass (Herbig Ae) stars continue to accrete material from their disks. The inner boundaries of these disks are expected to be at the co-rotation radius from the star, typically 3-5 stellar radii (~0.05 AU for the T Tauri stars). The environment closer to the star is controlled by a strong magnetic field, with accreting material channeled along field lines to the photosphere. In the accretion shock plasma temperatures increase from several thousand to a few million degrees. Due to the high temperatures, UV emission from the chromosphere and the accretion spot(s) is detectable at high contrast against the lower-temperature stellar photosphere[2]. While inner disk edges have been resolved by HST for dust disk cavities with radii in the 10-20 AU range[15] and IR disk images over a broader range have been obtained with ALMA[1], the inner edge of the gas disk has yet to be resolved for any young star with HST, but would be resolved with a *LBSI* for stars as distant as 160 pc. Fig. 3 shows a simulation of such an observation of the Ly$\alpha$-fluoresced $H_2$ emission originating in the inner portions of the dust disk of a T Tauri star at ~50 pc. Determining the size and geometry of the field-dominated region is of great importance for understanding stellar rotational braking, and accretion rates[7] as a function of global disk parameters. In addition to providing the size of the region, repeated observations may reveal rotation of resonances and indirectly point to the location of planets.

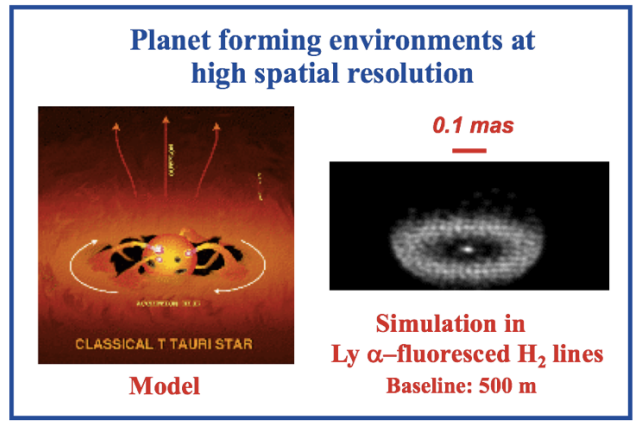

Fig. 3: Simulation of an observation of Ly$\alpha$-fluoresced $H_2$ emission from the inner disk regions of a T Tauri star at ~50 pc.

Direct spectropolarimetric observations of T Tauri stars[10,11] and the lack of X-ray eclipses[19] indicate that the accretion footprints on young stars can vary from pole to equator. Sub-mas spatial resolution will allow us to directly image the accretion hot spot(s), and provide a map of the accretion flow from the co-rotation radius of the disk onto the accretion footprints, using emission lines spanning a wide ionization range. Such imagery will allow us to test how the accretion geometry changes with stellar mass, age, and disk properties.

Young stars frequently drive bipolar outflows that can be traced up to parsec scales[25]. For nearby stars, these outflows can be resolved from the central star in H Ly$\alpha$ and optical forbidden lines, and in some cases in Si III 1206.5 Å[18]. HST imagery suggests that by a few AU from the star, the bipolar outflows are orthogonal to the disk. This is unexpected for outflows that are following the stellar magnetic field geometry, since the stellar-rotation and global-magnetic axes are not perfectly aligned in the Sun, or for the few PMS stars studied in detail[3,12]. One possibility is that the outflow launch and collimation region extends over the inner few AU of the disk[35], but the detailed coupling between mass loss from the star and the larger-scale jet is currently poorly understood. Sub-mas observations in the UV would also allow us to resolve any uncollimated wind component occurring from the star[13] or the disk[31]. The latter has been proposed as a means of transporting silicates and organics from the inner parts of the protoplanetary disk into distant icy planetesimals, thus accounting for the compositional diversity of comet nuclei[33].

***Dynamic Processes in Hot stars: Rotation, Disks, Winds, and Circumstellar Envelopes***
There are many competing processes on stars that produce structures on the surface or in the circumstellar environment. These processes include radiative winds, rapid rotation, pulsations, and magnetic fields, many of which may operate simultaneously within the stellar envelope.

*Understanding how massive stars rotate is important for the accurate modeling of stellar evolution and computing the final chemical yields of stars* [24]. Hot (O, B, Wolf-Rayet) stars tend to be the most rapidly rotating types of stars (excluding degenerate stars), and many are rotating so fast that their shapes are centrifugally distorted into oblate spheroids. Although rapid rotation in the very rare eclipsing binaries is measurable using light curves and radial velocity profiles, it is extremely difficult to pin down the detailed properties of single-star rapid rotation. A *LBSI* would enable direct measurement of the rotation rate and any differential rotation by imaging features moving across the star at different latitudes. Imaging the stellar oblateness will provide a better measure of the star's total angular momentum than feature-tracking alone could provide.

*Hot stars exhibit strong stellar winds that contribute significantly to the mass and energy balance of the interstellar medium*. Quantitative modeling of UV spectral features associated with stellar winds has evolved into a reasonably accurate means of deriving fundamental stellar parameters and distances [21]. The atmospheres and winds of hot stars are intrinsically variable, and it is now accepted that in many cases time-dependent phenomena (e.g., pulsations or magnetic field evolution) in the photosphere provide "shape and structure" to the wind [14]. The direct observational confirmation of a causal connection between specific stellar variations and specific wind variations, though, has proved elusive. For many O and B stars, it is not clear whether large-scale wind inhomogeneities are rotationally modulated (i.e., due to spots) or if pulsations are responsible, or if the variability occurs spontaneously in the wind. Sub-mas observations would shed light on the origins of wind variability. Simply seeing correlations between individual spots (no matter their physical origin) and modulations in the wind would be key to understanding how hot stars affect their local environments. One paradigm to be tested is the idea that discrete absorption components (DACs) are caused by corotating interaction regions (CIRs) in the winds[9]. While continuum-bandpass filters can be used efficiently to search for thermal and diffusive inhomogeneities on a hot star's disk, most other processes are best studied by imaging in UV spectral lines. From the ground one can do some imaging in Hα, but it is so optically thick that structures are hard to see. In the UV, however, the C IV doublet can be employed to study inner winds and co-orbiting structures of hot stars, while the Mg II doublet can be used to trace the discrete ejections of mass and the extent of disks out to several stellar radii.

Classical Be stars are ostensibly-single (though a binary fraction ~30% has been suggested [26]), rapidly-rotating, probably post-Zero Age Main Sequence (ZAMS) stars, which eject mass that episodically collapses onto an equatorial disk. The observed properties of Be stars and their circumstellar gas are consistent with the coexistence of a dense equatorial disk and a variable stellar wind[30]. A long-standing puzzle in hot-star astrophysics is the physical origin of this disk, both from the standpoint of mass supply (the winds may be too tenuous) and angular momentum and energy supply (the disk particles are in Keplerian orbits but the stellar surfaces are not), and the importance of binarity including the recently confirmed sdO companions (detected only in the FUV) that are the product of earlier mass and angular momentum transfer[28, 29, 38].

*Direct UV/Optical imaging of Be stars will provide answers regarding the physical distribution of matter, structures within the disks and winds (spiral density waves or clumpy structures), wind/disk interaction regions, and ionization structure, and the variation of these parameters with time*. Sub-mas resolution would allow characterization of the mean disk properties (e.g., inclination, radial density structure, thickness) with spectral type and rotation rate, which would provide empirical constraints on the wide variety of proposed, but unverified, theories.

Wolf-Rayet (WR) stars are believed to be the central, heavy cores of evolved O-type stars that have lost most of their hydrogen-rich outer layers as a stellar wind. *WR stars have observed mass loss rates at least an order of magnitude higher than other O stars (i.e., of order $10^{-4}$ $M_{sun}$/yr), and the origin of these extremely dense and optically thick outflows is still not well understood*. The only way that line-driven wind theory can account for such large mass loss rates is if the opacity in the lines is utilized many times (i.e., if photons multiply scatter through the optically thick outer atmosphere before they give up all of their radiative momentum to the gas[15]). However, other ideas exist, including fast magnetic rotation[20] and "strange-mode" pulsations in the chemically enriched interiors[17]. Direct imaging of winds would permit direct evaluation of these models, and an assessment of the role of turbulence[23] in these massive winds. Turbulence greatly complicates the study of line-driven mass loss[6] and makes it much more difficult to understand how WR winds are accelerated by radiation. Sub-mas UV observations are needed in order to identify the origin and nature of these clumps and clarify the physical processes responsible for these winds.

### *Pulsation Processes and their Impact on Stellar Structure and Mass Loss*

Pulsations are found in many different types of stars, ranging from very hot main-sequence stars to dying cool giants and supergiants, and stellar relics. *In many cases stellar pulsations, radial or non-radial, significantly affect the extent, composition, and structure of stellar atmospheres*. The signatures of pulsation are very prominent in the UV (e.g., Mg h & k lines) and a *LBSI* will enable direct imaging of pulsation effects including surface structures and shock fronts as they propagate through the dynamical atmospheres. Images of the effects of the pulsation will provide key inputs to hydrodynamical models for a range of diverse pulsators, such as Miras and Cepheids, cool supergiants, and hot B-stars. Direct observation of the shock-propagation in extended stellar atmospheres and winds will characterize the time evolution and spatial symmetries of shocks and constrain and improve theoretical shock models in stars with a wide range of masses. These observations will answer a large number of crucial questions about stellar interiors, core convection, chemical mixing, and magnetic fields.

Theories of non-radial pulsations are still evolving and the imaging of how rotation affects the latitudinal profile of pulsation amplitudes would verify or falsify certain modeling assumptions and directly diagnose the angular momentum profiles of these stars[36]. *For example, the direct imaging of a cause-and-effect relationship between stellar and circumstellar features could provide the long-sought explanation for the Be phenomenon*.

### *Convection in Cool Evolved Giant and Supergiant Stars*

Stars more than 1.5 $M_{sun}$ are not magnetically active during their mature life on the main sequence because they lack envelope convection. Consequently, they begin their transformation to red giant stars with essentially the same rotational energy they had after their formation. As they expand, a dynamo is activated once the star cools enough to develop envelope convection. That may lead to significant, sudden magnetic braking, which possibly results in a substantial difference between the rotation rates of the deep interior and the magnetically-active convective envelope[32]. Detailed understanding of the onset of dynamos in evolving stars with such shear layers between envelope and interior, and of the possible consequences for the internal dynamics, will greatly benefit from disk-resolved imaging and seismic observations.

Continuing their evolution as red giants, the stars reach a point where the coronal activity disappears again, to be replaced by substantial mass loss at much lower temperatures. Even though there is an absence of magnetically heated transition-region and coronal plasma in the late-K and M-type giant stars, their winds are thought to be driven by magneto-hydrodynamic waves. It has been proposed[31] that a coronal dividing line is a consequence of a dynamo transition from large-scale structures with closed field lines and coronal heating to small-scale structures with open field lines and increased mass loss. The hybrid stars that display both phenomena are the key to understanding the dividing line and the associated change in the dynamo mode from global to local. Sub-mas imaging of the transition-region and chromospheric emissions in the UV will reveal the magnetic field topology on stars on both sides of the dividing line, and on the hybrid-atmosphere stars.

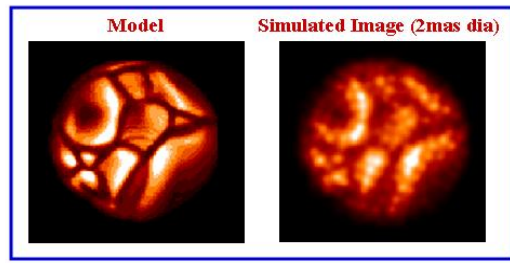

Stars with masses between ~10-30$M_{sun}$ eventually expand into red supergiants and the scale of the surface convection changes to the point that we expect only a few convective cells to cover the entire star, as observed for Betelgeuse[16]. Fig. 4 shows a model and simulated sub-mas observation of this convection. Recent observations of Pi$^1$ Gru with PIONIER[27] have confirmed the existence of these large cells and suggest the presence of large-scale turbulence, which may imply that a turbulent local dynamo may again create magnetic fields on a near-global scale. It has been demonstrated[4,37] that magnetic fields exist for these stars and are of the order of few (~1-2) G. A *LBSI* can image both the large-scale convection, its evolution, and the chromospheric patterns it drives.

Fig. 4: Model (Freytag) and simulated observation (500m baseline) of the convection on a supergiant (~ α Ori) at 2 kpc. These convective cells transport the energy from the interior to the surface, evolving on a timescale ~year, with ~dozen cells filling the surface.

*Evolution of Active Phenomena on Cool, Dwarf Stars and Impact on Exoplanets*
Sub-mas interferometry can also help us understand the evolution of active phenomena on cool dwarf stars. Even though such stars are small, many are close by with angular radii of a few milliarcsec. Spatial resolution of the winds of cool dwarf stars would be especially interesting to determine wind properties and mass-loss rates that are poorly known but important for understanding atmospheric loss from exoplanets, especially, for example, planets similar to Mars.

*Supernovae and Novae*
With the exception of the relatively nearby SN1987A, which could be well-studied by HST, it has not been possible to obtain much information about the close-in spatial structure of supernovae (typical sizes <1 mas). Radio VLBI observations have resolved a few supernovae, but are more a probe of the interaction of the SN shock front with the circumstellar material than of the supernova[5]. Narrowband HST images of novae[34] suggest they have very interesting 3-D structure (slow-moving equatorial rings plus bi-polar jets), and they are closer and brighter than the supernovae. Time-resolved spectra show that the ejecta are highly fragmented, and evolve significantly on timescales of days to weeks. Direct, sub-mas imaging would resolve early stages of expansion of supernovae at a few Mpc, and of galactic novae. These images would provide essential information on the nature of the explosion, especially in regard to its symmetry or asymmetry, and of the early evolution.